# Vortices in two-dimensional nanorings studied by means of the dynamical matrix method

S. Mamica

*Faculty of Physics, Adam Mickiewicz University in Poznań*
*ul. Umultowska 85, 61-614 Poznań, Poland*
E-mail: mamica@amu.edu.pl

This paper concerns an investigation of the spin wave excitations in magnetic nanoparticles. We provide a detailed derivation of the theoretical method for the determination of the normal modes of confined magnetic systems based on a discrete lattice of magnetic moments. The method is based on the damping-free Landau–Lifshitz equation and general enough to be utilized for the magnetic system of any dimensionality, magnetic structure, shape, and size. As an example we explore the influence of the competition between exchange and dipolar interactions on the spectrum of normal modes as well as on the stability of the vortex state in two-dimensional nanorings. We show the lowest-frequency mode to be indicative of the dipolar-to-exchange iterations ratio. We also study behavior of the fundamental mode and present the influence of both, the discreteness of the lattice and the dipolar-to-exchange iterations ratio, on its hybridization with azimuthal modes. We complete the paper with a selective review of the spin wave excitations in circular dots to compare with the results obtained for the rings.

## 1. Introduction

Coexistence of the short- and long-range interactions results in a variety of interesting effects. In thin magnetic films with natural surface it results in surface and subsurface localization [1–3], in patterned magnetic films is responsible for the splitting of the spectrum into subbands [4,5], and in magnonic crystals causes the complete bandgap to open [6–10]. In particular the competition between exchange and dipolar interactions has a significant influence on the spin-wave spectrum and leads to a variety of stable and metastable magnetic configurations. For example, in magnetic nanodots with a thickness small enough with respect to the diameter one of these configurations is the vortex state [11–15].

In the vortex configuration the in-plane components of the magnetic moments form a closed system (Fig. 1(a)). Circular dots and rings are circularly magnetized while in square rings magnetic moments form so called Landau state (a closure domain system) [16,17]. In square dots, as shown by simulations, the arrangement of the magnetic moments combines both Landau state near the borders and circular magnetization near the centre of the vortex [18]. This is a consequence of two competing effects. On the one hand, the center of the vortex tends towards the circular magnetization as a result of minimizing the (local) exchange interaction at the meeting point of four domains. On the other hand, the tendency to minimize the magnetic charges at the surface causes the magnetization to follow the edges of the dot. The circular magnetization appears in rather small area thus in large square dots the domain magnetization prevails in the major part of the dot, but in small dots only in minor corner regions the circular configuration fails to fit the geometry of the system.

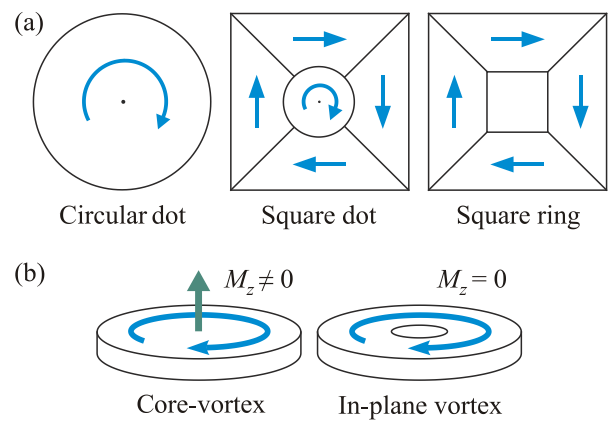

*Fig. 1.* (a) In-plane magnetization component in the vortex state for circular and square dot, and square ring. In the square dot there is a circular magnetization around the centre and Landau state near the edges. (b) Vortex with core vs in-plane vortex.

In full dots made of usual ferromagnetic materials like cobalt or permalloy exchange interactions are strong enough to form so called vortex core in the centre of the vortex (which does not necessarily coincide with the centre of the dot, though is close to it in stable vortices) [19–21]. The core of the vortex is a small region with nonzero out-of-plane component of static magnetization (Fig. 1(b)). In the rings the core does not appear because the central part of the dot is removed and thus the vortex centre lies outside of the system. In such case the static magnetization lies in the plane of the dot throughout its volume [22]. In fact, in full dots the in-plane (coreless) vortex can exists as well but for very weak exchange interactions [23].

Magnetic vortices are systems with the great potential applicability. They serve as microwave-frequency oscillators [24,25], can be used in frequency multiplication [26], or for data storage and information processing [27,28]. In all of these applications a crucial role is played by spin waves. They also have a significant influence on the stability of the magnetic configuration [29,30], even if the system is smaller than the characteristic exchange length [31]. Because of their importance spin wave excitations in magnetic dots have been widely studied in numerous papers experimentally [32–37], theoretically [38–43], and by means of numerical simulations [44–49]. On the other hand, the properties of spin waves in circularly magnetized rings are much less explored [50–52].

In the literature different types of modes are reported to be observed. For example, the observed lowest-frequency mode is an azimuthal mode of different order [53–56], a localized mode [45,57–59], or even the fundamental mode [37]. In Ref. 60 we use very effective method based on discretized version of Landau–Lifshitz equation to study this topic. We show that this diversity is related to the competition between the dipolar and exchange interactions. In the present paper we provide a detailed description of this method and show its usefulness with an example of magnetic rings, but the method is general enough to be applicable to any confined magnetic system regardless its dimensionality, internal structure, shape, or size.

The paper starts with a selective review of the properties of spin waves in circular dots with special attention paid on the lowest frequency mode (Sec. 2). In Sec. 3 we derive the theoretical method used in our approach and discuss its strong and weak points in comparison to the simulations. Next sections show our results obtained for magnetic rings: the spectrum of normal modes (Sec. 4), its dependence on the dipolar-to-exchange interaction ratio (Sec. 5), and the stability of the magnetic vortex vs the size and shape of the ring (Sec. 6). Finally, we close the paper with conclusions provided in Sec. 7.

**2. Spin wave excitations in circular dots**

In magnetic dots in the core-vortex state two types of excitations can be distinguished, namely gyrotropic motion of the core and spin waves [61,57]. The gyrotropic mode is low-frequency excitation (in the range of hundreds of MHz) related to the precession of the vortex core as a whole in the effective magnetic field. Whereas spin waves are stationary excitations (normal modes) of the system of magnetic moments. Depending on the size of a dot their frequencies start at several GHz. In circular dots their profiles are similar to those of vibrations of the circular membrane and the same labeling $(k,m)$ is useful to describe them, where the radial number $k$ is the number of nodal lines in the radial direction and the azimuthal number $m$ in azimuthal direction.

The occurrence of the nodal line means a phase shift along particular direction (radial or azimuthal). We will call azimuthal modes the spin waves which have radial number $k = 0$, i.e. with no phase shift as we move from the dot centre to its boundary. In the literature there is some inconsistency in the labeling of such modes. In some papers they are labeled as $k = 0$ [33,35,57] while in others as $k = 1$ [53–55] (probably because of the vanishing amplitude of oscillation at the dot boundary or the vortex core). This inconsistency is not of appreciable importance unless the results are compared with analytical models and even in this case the most important thing is to remember that it exists in the literature.

The results obtained in experimental and micromagnetic studies depend significantly on the method of the spin waves excitation and the observation technique. For example, if spin waves are excited by a short pulse of magnetic field the observed spectrum depends on the orientation of the pulse with respect to the plane of the dot. (A similar dependence has been observed in micromagnetic simulations.) For the pulse oriented in the dot plane (in-plane pumping) only modes with non-zero azimuthal number $m$ are excited (they can have non-zero radial number as well) while for the pulse perpendicular to the dot (perpendicular pumping) the excited modes are $m = 0$ modes [37,54–57]. In the last case the observed lowest-frequency mode is usually fundamental mode (0,0) [37,54–56]. On the other hand, for in-plane pumping a first order azimuthal mode is usually reported as the lowest one [33,35,54,55,57].

An exception is reported by Zhu et al. [57]. For perpendicular pumping they observe an extra mode with a much lower frequency than fundamental one, however, higher that the gyrotropic mode. This mode does not occur in the micromagnetic simulations presented in [57] but the authors suppose that this could be edge-localized mode. It is also absent for in-plane pumping which suggests strong out-of-plane component of the dynamical magnetization. In our theoretical investigation we found the edge modes usually tend to occur higher in the spectrum [60], although the centre-localized mode appears as the lowest if it become soft-mode, i.e. for metastable magnetic configuration. In the case of strong exchange interactions it oscillates almost purely out-of-plane. Rivkin et al. [42]. found theoretically the lowest mode to be localized at the centre of the dot (the vortex core), but with non-negligible both in-plane and out-of-plane oscillations. In micromagnetic simulations the lowest mode was found to be localized at the vortex core by Boust et al. [58]. In this case the oscillations are purely in-plane which, together with the low frequency, might suggest a gyrotropic mode. On the other hand, according to our own results, spin waves with similar profile occur in in-plane vortices for strong dipolar interactions (weak exchange). Also the edge anisotropy could reduce the frequencies of the localized modes. The edge-localised mode is reported to be the lowest by Novosad et al. [45]. and by Liu et al. [59].

The lowest-frequency mode has been identified as an azimuthal mode of different order by Buess et al., even if they use the perpendicular pulse [53]. A $(0,2)$ mode was found to be the lowest in dots of 4 μm and 6 μm in a diameter and 15 nm thick. In smaller dots (2 and 3 μm in diameter) the $(0,1)$ mode is reported to be the lowest one. This experimental result supports theoretical studies provided by Ivanov and Zaspel [40], by Ziveri and Nizzoli [62] and by ourselves [60]. In papers [40] and [62] using analytical methods authors show that in dots of large enough diameter second order azimuthal mode has the lowest frequency while for smaller diameter first order one. Our own results [60] also agree with this statement: higher-order azimuthal modes descend the spectrum as an effect of increasing the diameter of the dot. Moreover, we observe the same effect with increasing dipolar-to-exchange interaction ratio. The general conclusion is that in circular dots in the vortex state the competition between the dipolar and exchange interactions is reflected by the character of the lowest-frequency spin-wave excitation. The exchange interactions favour $m=1$ modes while the dipolar interaction higher-order azimuthal modes, regardless of whether their predomination is due to the size or material of the dot.

If dipolar interactions are strong enough the above rule leads to the negative dispersion relation, i.e. decreasing frequency with increasing azimuthal number. Such effect was found analytically in already cited papers [40] and [62] as well as experimentally in [53].

## 3. Dynamical matrix method

The system being the subject of this study is a circular ring cut out from a 2D discrete lattice with elementary magnetic moments (spins) in lattice sites. An example of the ring based on the square lattice is shown in Fig. 2. Obviously, the boundaries of such ring are not perfectly circular and by "circular" we understand a system cut out by means of circles. The system is naturally discrete therefore its edges cannot be smoothed (as it is in continuous systems with artificial discretization, e.g., in micromagnetic simulations) [63]. Of course, the edge smoothness increases with the size of the ring (measured in lattice constants).

The dynamics of a single magnetic moment $\mathbf{M_R}$, $\mathbf{R}$ being the position vector, is considered in the linear approximation, assuming $|\mathbf{m_R}| \ll |\mathbf{M_R}|$, $|\mathbf{M_{0,R}}| \simeq |\mathbf{M_R}|$ and $\mathbf{m_R} \perp \mathbf{M_{0,R}}$, where $\mathbf{M_{0,R}}$ and $\mathbf{m_R}$ are the static and dynamic component of the magnetic moment, respectively. The static component defines one of the axis of a local Cartesian coordinate system, $\mathbf{i_R}$ (see Fig. 2). Second unit vector, $\mathbf{j_R}$, lies in the plane of the dot and is oriented towards its centre. In the case of an in-plane vortex studied here, the third basis vector, $\mathbf{k_R}$, is perpendicular to the dot plane at each lattice site. Thus, the dynamic component of the magnetic moment is given by $\mathbf{m_R} = m_{j,\mathbf{R}} \mathbf{j_R} + m_{k,\mathbf{R}} \mathbf{k_R}$, where $m_{j,\mathbf{R}}$ and $m_{k,\mathbf{R}}$ are the in-plane and out-of-plane coordinates, respectively.

*Equations of motion*

To describe the time evolution of $\mathbf{M_R}$ we use the discrete damping-free Landau–Lifshitz (LL) equation:

$$\frac{\partial \mathbf{M_R}}{\partial t} = \gamma \mu_0 \mathbf{M_R} \times \mathbf{H_R^{eff}}, \qquad (1)$$

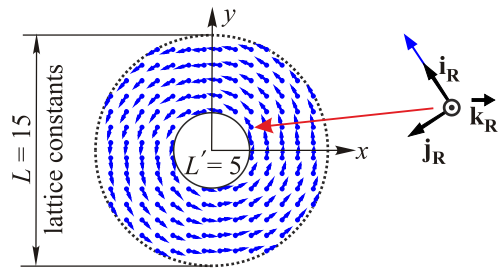

*Fig. 2.* (Color online) Schematic plot of a circularly magnetized ring based on a 2D square lattice. Magnetic moments (represented by the arrows) are arranged in the lattice sites within the ring. The external size $L$ is defined as the number of lattice constants in the diameter of the circle used for cutting out the ring (the dotted circle). Similarly, the internal size $L'$ is the radius of the inner circle (in units of the lattice constant). To the right, the local coordinate system associated with the magnetic moment indicated by the arrow.

where $\gamma$ is the gyromagnetic ratio, $\mu_0$ is the vacuum permeability, and $\mathbf{H}_\mathbf{R}^{\text{eff}}$ the effective magnetic field acting on the magnetic moment at the position $\mathbf{R}$. In the linearized form the LL equation reads:

$$i\frac{\omega}{\gamma\mu_0}\mathbf{m}_\mathbf{R} = \mathbf{M}_{0,\mathbf{R}} \times \mathbf{h}_\mathbf{R} + \mathbf{m}_\mathbf{R} \times \mathbf{H}_\mathbf{R}, \quad (2)$$

where $i$ is the imaginary unit, $\omega$ is the frequency of harmonic oscillation of $\mathbf{m}_\mathbf{R}$, $\mathbf{H}_\mathbf{R}$ and $\mathbf{h}_\mathbf{R}$ are static and dynamic components of the effective field, respectively. In our model we will include four terms of the effective field: external field, dipolar and exchange interactions, as well as uniaxial anisotropy.

Assuming the external field to be uniform and time-independent it has only static component: $\mathbf{H}_0 = \mathbf{B}_0/\mu_0$.

The dipolar field acting on a lattice site $\mathbf{R}$ is given by:

$$\mathbf{H}_\mathbf{R}^d = \frac{1}{4\pi a^3} \sum_{\mathbf{R}'\neq\mathbf{R}} \left( \frac{3(\mathbf{R}'-\mathbf{R})(\mathbf{M}_{\mathbf{R}'}\cdot(\mathbf{R}'-\mathbf{R}))}{|\mathbf{R}'-\mathbf{R}|^5} - \frac{\mathbf{M}_{\mathbf{R}'}}{|\mathbf{R}'-\mathbf{R}|^3} \right),$$

where $a$ is the lattice constant and the summation $\sum_{\mathbf{R}'\neq\mathbf{R}}$ runs over all the lattice sites except that indicated by $\mathbf{R}$. The position vectors are expressed in the units of $a$.

Assuming the exchange interaction uniform and limited to nearest neighbors (NN) only, the exchange field is given by the formula:

$$\mathbf{H}_\mathbf{R}^{\text{ex}} = \frac{2J}{\mu_0(g\mu_B)^2} \sum_{\mathbf{R}'\in NN} \mathbf{M}_{\mathbf{R}'},$$

$J$ being the exchange integral. The summation runs over all $\mathbf{R}'$ being the position vectors of NN of the spin indicated by $\mathbf{R}$.

The uniaxial anisotropy field reads:

$$\mathbf{H}_\mathbf{R}^a = \frac{2D_\mathbf{R}}{\mu_0(g\mu_B)^2}(\mathbf{M}_\mathbf{R}\cdot\mathbf{K}_\mathbf{R})\mathbf{K}_\mathbf{R},$$

where $\mathbf{K}_\mathbf{R}$ is the unit vector in the direction of the anisotropy easy axis and $D_\mathbf{R}$ is the anisotropy coefficient, both in position $\mathbf{R}$.

Assuming that all spins $S$ within the dot are the same the length of the static component of the magnetic moment is independent on the position in the dot, so $\mathbf{M}_{0,\mathbf{R}} = g\mu_B S \mathbf{i}_\mathbf{R}$. Thus the components of the effective field can be written as:

$$\mathbf{H}_\mathbf{R} = \frac{2S}{g\mu_B\mu_0}\tilde{\mathbf{H}}_\mathbf{R} \quad\text{and}\quad \mathbf{h}_\mathbf{R} = \frac{2}{(g\mu_B)^2\mu_0}\tilde{\mathbf{h}}_\mathbf{R}, \quad (3)$$

where

$$\tilde{\mathbf{H}}_\mathbf{R} = \tilde{b}\mathbf{B}_0 + J\sum_{\mathbf{R}'\in NN}\mathbf{i}_{\mathbf{R}'} + D(\mathbf{i}_\mathbf{R}\cdot\mathbf{K}_\mathbf{R})\mathbf{K}_\mathbf{R} +$$

$$+ \tilde{d}\sum_{\mathbf{R}'\neq\mathbf{R}}\left(\frac{3(\mathbf{R}'-\mathbf{R})(\mathbf{i}_{\mathbf{R}'}\cdot(\mathbf{R}'-\mathbf{R}))}{|\mathbf{R}'-\mathbf{R}|^5} - \frac{\mathbf{i}_{\mathbf{R}'}}{|\mathbf{R}'-\mathbf{R}|^3}\right) \quad (4)$$

and

$$\tilde{\mathbf{h}}_\mathbf{R} = J\sum_{\mathbf{R}'\in NN}\mathbf{m}_{\mathbf{R}'} + D(\mathbf{m}_\mathbf{R}\cdot\mathbf{K}_\mathbf{R})\mathbf{K}_\mathbf{R} +$$

$$+ \tilde{d}\sum_{\mathbf{R}'\neq\mathbf{R}}\left(\frac{3(\mathbf{R}'-\mathbf{R})(\mathbf{m}_{\mathbf{R}'}\cdot(\mathbf{R}'-\mathbf{R}))}{|\mathbf{R}'-\mathbf{R}|^5} - \frac{\mathbf{m}_{\mathbf{R}'}}{|\mathbf{R}'-\mathbf{R}|^3}\right). \quad (5)$$

The coefficients are: the external field coefficient $\tilde{b} = g\mu_B/2S$ and the dipolar coefficient $\tilde{d} = (g\mu_B)^2 \times \mu_0/8\pi a^3$.

Dividing (2) by $2S/g\mu_B\mu_0$ and introducing $\Omega = g\mu_B\omega/2\gamma S$ one can obtain the equation of motion in the compact form:

$$i\Omega\mathbf{m}_\mathbf{R} = \mathbf{i}_\mathbf{R}\times\tilde{\mathbf{h}}_\mathbf{R} + \mathbf{m}_\mathbf{R}\times\tilde{\mathbf{H}}_\mathbf{R}.$$

This vector equation can be rewritten as a set of equations for three coordinates of the magnetic moment:

$$0 = m_{j,\mathbf{R}}\tilde{H}_{k,\mathbf{R}} + m_{k,\mathbf{R}}\tilde{H}_{j,\mathbf{R}},$$

$$i\Omega m_{j,\mathbf{R}} = -\tilde{h}_{k,\mathbf{R}} + m_{k,\mathbf{R}}\tilde{H}_{i,\mathbf{R}},$$

$$i\Omega m_{k,\mathbf{R}} = \tilde{h}_{j,\mathbf{R}} - m_{j,\mathbf{R}}\tilde{H}_{i,\mathbf{R}},$$

where the subscripts $i$, $j$, and $k$ denote respective coordinates of the vectors (in local coordinate system).

Finally, the system of equations of motion for the dynamic components of magnetic moments is following:

$$i\Omega m_{j,\mathbf{R}} = m_{k,\mathbf{R}}\left[\tilde{b}\mathbf{B}_0\cdot\mathbf{i}_\mathbf{R} + J\sum_{\mathbf{R}'\in NN}\mathbf{i}_\mathbf{R}\cdot\mathbf{i}_{\mathbf{R}'} + D_\mathbf{R}(\mathbf{i}_\mathbf{R}\cdot\mathbf{K}_\mathbf{R})^2 + \right.$$

$$\left. + \tilde{d}\sum_{\mathbf{R}'\neq\mathbf{R}}\left(\frac{3[(\mathbf{R}'-\mathbf{R})\cdot\mathbf{i}_\mathbf{R}][(\mathbf{R}'-\mathbf{R})\cdot\mathbf{i}_{\mathbf{R}'}]}{|\mathbf{R}'-\mathbf{R}|^5} - \frac{\mathbf{i}_\mathbf{R}\cdot\mathbf{i}_{\mathbf{R}'}}{|\mathbf{R}'-\mathbf{R}|^3}\right)\right] -$$

$$- J\sum_{\mathbf{R}'\in NN}\mathbf{k}_\mathbf{R}\cdot\mathbf{m}_{\mathbf{R}'} - D_\mathbf{R}(\mathbf{m}_\mathbf{R}\cdot\mathbf{K}_\mathbf{R})(\mathbf{k}_\mathbf{R}\cdot\mathbf{K}_\mathbf{R}) -$$

$$- \tilde{d}\sum_{\mathbf{R}'\neq\mathbf{R}}\left(\frac{3[(\mathbf{R}'-\mathbf{R})\cdot\mathbf{k}_\mathbf{R}][(\mathbf{R}'-\mathbf{R})\cdot\mathbf{m}_{\mathbf{R}'}]}{|\mathbf{R}'-\mathbf{R}|^5} - \frac{\mathbf{k}_\mathbf{R}\cdot\mathbf{m}_{\mathbf{R}'}}{|\mathbf{R}'-\mathbf{R}|^3}\right),$$

$$i\Omega m_{k,\mathbf{R}} = -m_{j,\mathbf{R}}\left[\tilde{b}\mathbf{B}_0\cdot\mathbf{i}_\mathbf{R} + J\sum_{\mathbf{R}'\in NN}\mathbf{i}_\mathbf{R}\cdot\mathbf{i}_{\mathbf{R}'} + D_\mathbf{R}(\mathbf{i}_\mathbf{R}\cdot\mathbf{K}_\mathbf{R})^2 + \right.$$

$$\left. + \tilde{d}\sum_{\mathbf{R}'\neq\mathbf{R}}\left(\frac{3[(\mathbf{R}'-\mathbf{R})\cdot\mathbf{i}_\mathbf{R}][(\mathbf{R}'-\mathbf{R})\cdot\mathbf{i}_{\mathbf{R}'}]}{|\mathbf{R}'-\mathbf{R}|^5} - \frac{\mathbf{i}_\mathbf{R}\cdot\mathbf{i}_{\mathbf{R}'}}{|\mathbf{R}'-\mathbf{R}|^3}\right)\right] +$$

$$+ J\sum_{\mathbf{R}'\in NN}\mathbf{j}_\mathbf{R}\cdot\mathbf{m}_{\mathbf{R}'} + D_\mathbf{R}(\mathbf{m}_\mathbf{R}\cdot\mathbf{K}_\mathbf{R})(\mathbf{j}_\mathbf{R}\cdot\mathbf{K}_\mathbf{R}) +$$

$$+ \tilde{d}\sum_{\mathbf{R}'\neq\mathbf{R}}\left(\frac{3[(\mathbf{R}'-\mathbf{R})\cdot\mathbf{j}_\mathbf{R}][(\mathbf{R}'-\mathbf{R})\cdot\mathbf{m}_{\mathbf{R}'}]}{|\mathbf{R}'-\mathbf{R}|^5} - \frac{\mathbf{j}_\mathbf{R}\cdot\mathbf{m}_{\mathbf{R}'}}{|\mathbf{R}'-\mathbf{R}|^3}\right).$$

$$(6)$$

*Dynamical matrix*

The system of Eqs. (6) can be represented as a following eigenvalue problem:

$$\Omega \begin{bmatrix} m_j \\ m_k \end{bmatrix} = \mathcal{M} \begin{bmatrix} m_j \\ m_k \end{bmatrix}, \quad (7)$$

where $m_j$ and $m_k$ are column vectors containing in-plane and out-of-plane amplitudes of magnetic moment precession, respectively, for every lattice site within the system.

Numerical diagonalization of the matrix $\mathcal{M}$, which is called dynamical matrix, yields the reduced frequencies $\Omega$ of the spin-wave excitations as well as their profiles, i.e. the distribution of amplitudes of the magnetic moments precession $m_j$ and $m_k$. It is worth to notice that derivation of Eqs. (6) is general enough to apply the model to the magnetic periodic system of any dimensionality, internal structure, shape, size or configuration of magnetic moments. For example in Ref. 18 we use this model to explore the spin wave profiles for core-vortices in square dots with the magnetic configuration obtained from simulations. Moreover, neglecting the exchange interaction made the model useful also for nonperiodic structures. Another advantage of the dynamical matrix method is that its diagonalization yields directly the spin-wave frequencies and profiles, without recourse to the Fourier transformation used in time-domain simulations.

In our approach we do not involve any simulations; the magnetic configuration is assumed concerning the size and shape of the system along with the interactions taken into account. Such assumed configuration may be unstable, but in this case in the spin-wave spectrum will occur nucleation modes with frequency equal zero which are responsible for reconfiguration of the magnetization. In other words, the lack of zero-frequency modes is indicative of the stability (or metastability) of the assumed magnetic configuration [64]. In comparison with the simulation methods this approach allows quick exploration of various configurations, the stability of which is inferred from the obtained spin-wave spectrum.

The assumption, rather than simulation, of the magnetic configuration is the main disadvantage of the method. Thus this approach can only be used for studying relatively simple magnetic configurations involving as few assumptions as possible. For example, the study of a core vortex will require additional assumptions regarding the shape and size of the core. In such cases a combined method seems to be the best solution, with simulations used for finding the stable configuration, and the dynamical matrix technique for calculating the corresponding spin-wave modes. On the other hand, for simple magnetic configurations such in-plane vortex in circular dots our results are in perfect agreement with simulations (compare, e.g., Refs. 23, 25).

The time dependence of the dynamical component is given by $\mathbf{m}(t) = \mathbf{m} \exp(i\omega t)$. Eigenvectors obtained from diagonalization of the dynamical matrix $\mathcal{M}$ are complex and the phase shift between the real (Re) and imaginary part (Im), i.e. $\pi/2$, gives $T/4$ shift in time, where $T = 2\pi/\omega$ is a period of oscillations for a given mode. Thus for the given time $t$ we have:

$$m_j(t) = [\text{Re}(m_j) + i\text{Im}(m_j)] \exp(-i\omega t) + \\ + [\text{Re}(m_j) - i\,\text{Im}(m_j)] \exp(i\omega t),$$

$$m_k(t) = [\text{Re}(m_k) + i\,\text{Im}(m_k)] \exp(-i\omega t) + \\ + [\text{Re}(m_k) - i\,\text{Im}(m_k)] \exp(i\omega t).$$

In other words, for time $t = 0$ all dynamic components are given by their real parts but after $t = T/4$ by imaginary parts.

*Dipolar-exchange case*

The vortex magnetic configuration is a result of the competition of dipolar and exchange interactions. Thus in the next part of our study we will restrict ourselves to these two terms in the effective field assuming no external field nor uniaxial anisotropy. In this case dividing resulting equations by $J$ changes the definition of reduced frequency and dipolar coefficient to the following:

$$\Omega = \frac{g\mu_B}{2\gamma SJ}\omega \quad (8)$$

and

$$d = \frac{(g\mu_B)^2 \mu_0}{8\pi a^3 J}. \quad (9)$$

New coefficient $d$ is the dipolar-to-exchange interaction ratio and this is the only material parameter in the dipolar-exchange case of our model. In typical ferromagnets strong exchange interaction results in very low values of $d$. For example, in the SPEELS experiments reported in Ref. 66 the exchange integral in an ultrathin film of Co on a Cu(001) substrate is estimated at 15 meV. From the definition (8) we get $d = 0.00043$.

### 4. Normal modes in circular rings

Let us consider the circular ring of size $L = 63$ and $L' = 32$ cut from the square lattice. Such ring consists of 2284 spins. Assuming an in-plane vortex (circular magnetization) as a magnetic configuration and the dipolar-to-exchange interaction ratio $d = 0.022$ we obtain the spin-wave frequency spectrum shown in Fig. 3. There are no zero-frequency modes in the spectrum which, according to the above discussion, is indicative of the (meta)stability of assumed magnetic configuration. The overall shape of the spectrum is typical for dipolar-exchange systems: for low mode numbers the shape is determined by dipolar interac-tions while by exchange interactions for high mode numbers. Another interesting feature is that some modes degenerate in pairs (see the inset).

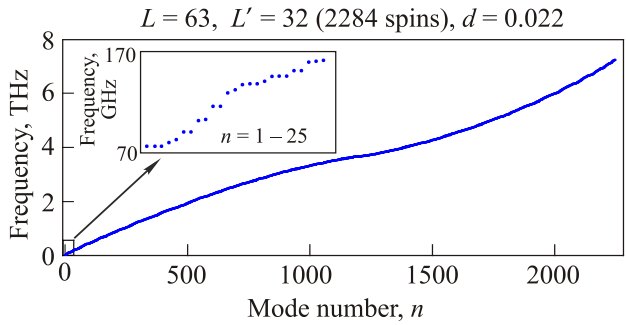

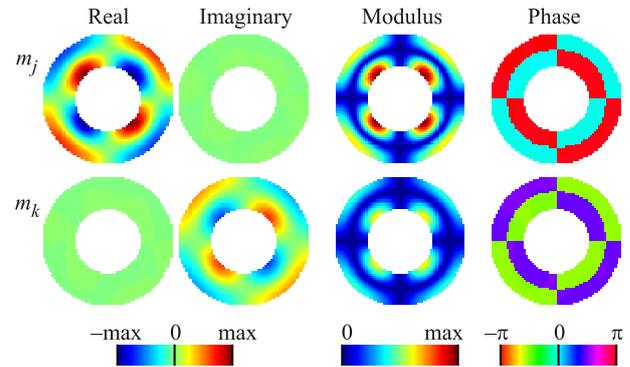

*Fig. 3.* Spin wave spectrum for a 2D circularly magnetized ring of size $L = 63$ and $L' = 32$ (2284 spins). Dipolar-to-exchange interaction ratio $d$ is set to 0.022. In the inset lowest 25 modes are shown. The lack of zero-frequency modes is indicative of the stability of the assumed magnetic configuration.

*Fig. 4.* (Color online) Different ways of presentation of a spin wave profile: real and imaginary part vs modulus and phase of the distribution of the in-plane ($m_j$) and out-of-plane ($m_k$) amplitude of precession of magnetic moments. An exemplar profile calculated for a ring of size $L = 63$ and $L' = 32$ (2284 spins) in the in-plane vortex state.

Before we go to the analysis of the spin wave profiles lets look at different ways of their presentation shown in Fig. 4. Since both dynamic coordinates of the magnetic moment, namely in-plane and out-of-plane, are complex they can be expressed as real and imaginary part or as modulus and phase. As we can see for both components the profiles are similar. Moreover, one part (real or imaginary) is sufficient to fully describe the distribution of the oscillation amplitude. Therefore in the later if the situation is clear we show only one part (Re or Im) of the one component (in-plane or out-of-plane). To describe the type of the mode we will use following notation: $(k, m)$, where the radial number $k$ and the azimuthal number $m$ specify the number of nodal lines in the respective directions. In this notation the exemplar mode in Fig. 4 is radial mode $(1, 2)$.

In Fig. 5 we show spin-wave profiles of some selected modes from the spectrum shown in Fig. 3. These are 18 lowest modes and two additional modes 43rd and 46th. Between the lowest modes the majority are azimuthal modes, i.e. the modes with no nodal lines in the radial direction. In this group modes come in pairs having the same absolute value of the azimuthal number. (For our purpose the sign of the azimuthal number is of no importance and the azimuthal number should be understood as $|m|$.) Some of these pairs are degenerated with both modes having the same frequency. In fact, this is quite typical and the question arises why for some pairs degeneration is lifted. The pairs of modes with an even azimuthal number, i.e. the modes $(0,2)$, $(0,4)$ and $(0,6)$, are not degenerate. This fact is related to their symmetry, the same as that of the lattice from which the dot has been cut out. For example, one of the $(0,2)$ modes ($n = 4$) has nodal lines along the high spin density lines ($x$ and $y$ axis in Fig. 2). In contrast, in the other $(0,2)$ mode ($n = 5$) the high spin density lines coincide with anti-nodal lines. An analogical situation occurs in periodic structures, in which a band gap forms between two states at the boundary of the Brillouin zone if one state has nodes and the other anti-nodes in the potential wells.

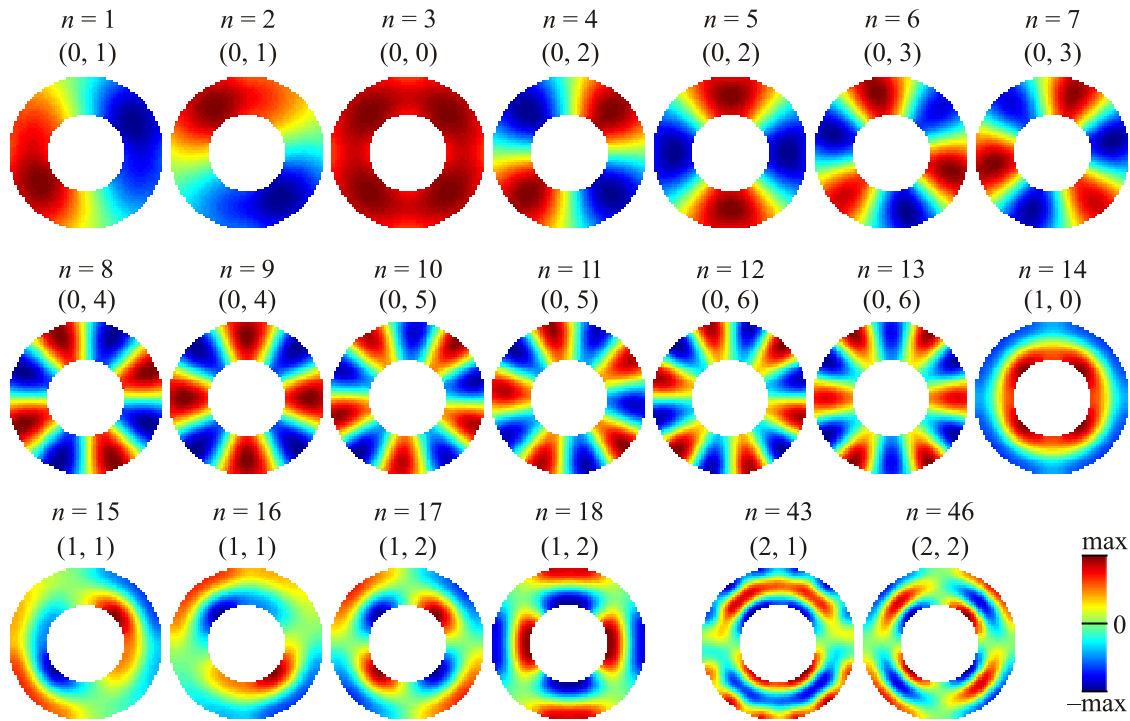

*Fig. 5.* (Color online) The spin-wave profiles of selected modes for a ring of size $L = 63$ and $L' = 32$ in the in-plane vortex state for $d = 0.022$ (the same ring as in Fig. 3). In brackets radial and azimuthal numbers are given; modes are also labelled with their mode number $n$.

In the literature there are reports of lifted degeneracy of azimuthal modes in core vortices due to the coupling of the spin waves with the gyrotropic motion of the core [67,68]. Since in the ring the coreless vortices exist, coupling with the motion of the core is out of the question. The nondegeneracy is due to the fact that the dot has been cut out from a discrete lattice and it is the symmetry of the lattice that determines which modes have lifted degeneracy. For example in the case of hexagonal lattice the frequency splitting occurs for the modes with azimuthal number divisible by 3 (see our paper [69] concerning hexagonal lattice).

Next group are radial modes, the modes with non-zero radial number. Modes 14th to 18th are first order radial modes with increasing azimuthal number. Similarly to azimuthal modes pairs of radial modes with even azimuthal number have lifted degeneracy while these with odd $m$ are degenerated. Additionally we show two examples of second order radial modes: $(2,1)$ (43rd mode) and $(2,2)$ (46th mode).

Within every group of modes with the same radial number we observe positive dispersion, i.e. the azimuthal number increases with increasing mode number (whit an exception of $(0,0)$ mode, $n = 3$). As we discuss in Sec. 2 this is indicative of predominant exchange interaction.

The 3rd mode presented in Fig. 5 ($n = 3$) is so called fundamental mode, which is the analogue of the uniform excitation. Its profile is not perfectly uniform but there are no nodal lines in any direction, azimuthal nor radial, so the mode is labeled as $(0,0)$. This nonuniformity is inherited form the symmetry of the lattice from which the dot has been cut out and results from the fact that the circular symmetry of the dot is broken by the symmetry of the discrete lattice. In our model the lattice discreteness is an intrinsic feature of the system and the nonuniformity of the fundamental mode profile is its natural consequence. Quite similar effect has been observed in micromagnetic simulations [56,70,71], in which, however, it stems from the artificial discretization of a continuous system into cubes (e.g. in the very popular OOMMF simulations) and this effect should not occur if the discretization is based on tetrahedral, e.g. in the NMAG approach [72].

## 5. Influence of dipolar interactions

In Sec. 4 we discuss the spin waves spectrum for particular value of the dipolar-to-exchange interaction ratio $d$. In Fig. 6 we show the dependence of the spectrum vs $d$ for the ring concerned previously ($L = 63$, $L' = 32$, 2284 spins). In the figure only 40 lowest modes are plotted and the frequency range is cut at 120 GHz. The dependence clearly indicates the existence of three ranges of $d$. In two of them, i.e. below $d_1$ and above $d_2$, the frequency of the lowest mode is zero which means the assumed magnetic configuration in unstable. On the other hand, for $d_1 < d < d_2$ the absence of the zero-frequency modes indicates that the in-plane vortex is (meta)stable. This picture reflects the nature of the magnetic in-plane vortex, which appears as a compromise between the exchange and dipolar interactions. When the exchange interaction is strong enough ($d$ is too low) they lead to the exchange-driven reorientation usually evidenced by appearing of the out-of-plane component of the static magnetic moment. If the exchange interaction is too weak ($d$ is too large) the dipolar interaction leads to a multi-domain or multi-vortex state and a dipolar-driven reorientation occurs. The stability of the in-plane vortex vs the size and the shape of the system is discussed in Sec. 6.

In the in-plane vortex stability regime for the majority of modes the frequency decreases significantly with decreasing exchange interaction (increasing $d$). These modes can be divided into two classes: starting from the $(0,1)$ mode, the rate of frequency decrease grows with increasing azimuthal number. However, above 80 GHz this rate is sensibly lower for some modes. These modes have a low azimuthal number and the radial number 1. In other words, the situation repeats: starting from the (1,1) mode, for which the rate of frequency decrease is almost the same as for the (0,1) mode, the frequency decrease rate grows with increasing $m$. Thus, the influence of the dipolar-to-exchange interaction ratio on the frequency of a mode is mainly determined by its azimuthal number, the radial number being of little impact.

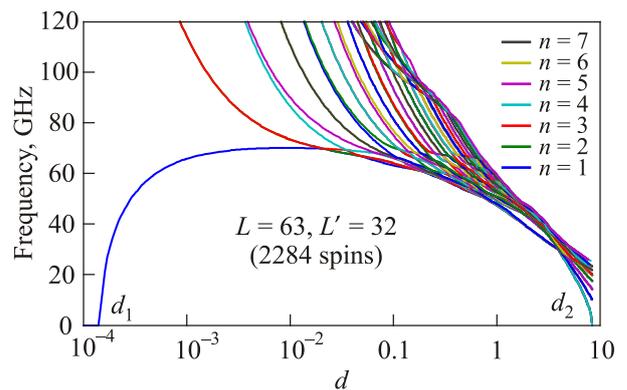

*Fig. 6.* (Color online) The frequency of 40 lowest modes as a function of the dipolar-to-exchange interaction ratio $d$ (in logarithmic scale) for a 2D circular ring of size $L = 63$ and $L' = 32$. The lack of zero-frequency modes is indicative of the stability of the assumed magnetic configuration (an in-plane vortex) for $d_1 < d < d_2$. The color assignment of the first seven mode lines is indicated at the right; the colors repeat cyclically for successive modes.

*Lowest mode evolution*

Since the frequencies of different spin-wave modes decrease at different rates with increasing $d$ the order of modes in the spectrum changes dynamically. In particular, the character of the lowest-frequency mode changes as a result of mode crossing. In Fig. 7 we show the spin wave profiles of the lowest frequency mode for different values of $d$. For $d$ close to $d_1$ the lowest mode is a soft mode; its frequency reaches zero for $d = d_1$, and the mode becomes a nucleation mode, responsible for the magnetic reconfiguration of the system. In the case considered this mode is the fundamental mode with a nearly uniform profile of the dynamic component of magnetization ($d = 0.005$ in Fig. 7). After mode crossing for $d \approx 0.02$ two degenerated azimuthal modes (0,1) are the lowest in the spectrum ($d = 0.035$) until second mode crossing at about 0.048, when one of the second order azimuthal modes become the lowest ($d = 0.1$). Next mode crossings cause higher order azimuthal modes to have the lowest frequency at $d \approx 0.164$, 0.192, 0.288 and 0.324 (see mode profiles for $d = 0.18$, 0.25, 0.3 and 0.5 in Fig. 7). Another feature is the tendency of the profile to concentrate the high amplitude in smaller area of the ring while $d$ increases (compare the profiles for azimuthal number $m = 2-6$ presented in Figs. 5, 7). Finally, its results in strong concentration of the profile around the lines of the highest spin density lines for $d > 1$ ($d = 4.0$ in Fig. 7).

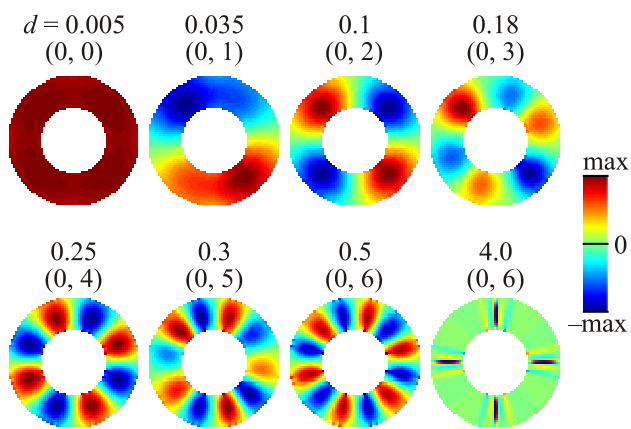

*Fig. 7.* (Color online) Spin-wave profiles of the lowest mode for different values of $d$ for the same ring as in Fig. 6.

**Fundamental mode evolution**

When the fundamental mode is no longer soft mode ($d > 0.001$) its frequency start to be only little dependent on $d$ (Fig. 8(a)). As we have already mentioned this weak dependence on $d$ results from its profile which is not perfectly uniform. The non-uniformity is inherited from the discreteness of the lattice thus the profile has the same symmetry (Fig. 8(b), $d = 0.05$). While azimuthal modes frequencies decrease with growing $d$ the fundamental mode ascend the spectrum due to the modes crossings until it meets the azimuthal mode of the same symmetry. For $d$ in the range 0.07–0.10 fundamental mode frequency significantly depends on $d$ due to the repulsion with one of the (0, 4) modes. This is the result of the mode hybridization shown in Fig. 8(c) where the profiles of three modes are shown: fundamental one and both fourth order azimuthal (marked with black dots in Fig. 8(a)). The hybridization is a consequence of the symmetry of hybridizing modes: both profiles, for (0, 0) and (0, 4) modes, have four-fold symmetry. On the other hand, only one (0, 4) mode ($n = 5$) is involved in the hybridization — its profile is mixed with the profile of the fundamental mode. The second (0, 4), $n = 8$, is not affected by the vicinity of the fundamental one. This means there is the second condition for the hybridization: the matching of the in-phase anti-nodes in the azimuthal mode with the maximums in the fundamental mode. Only one of the two (0, 4) modes, namely that with anti-nodal lines in the same phase coinciding with the amplitude maximums in the fundamental mode, is involved in the hybridization, and consequently repulsion. The other (0, 4) mode is "ignored" by the fundamental mode, in spite of the same symmetry of the two. This is because the amplitude maximums of the fundamental mode match the nodal lines in the profile of this (0, 4) mode.

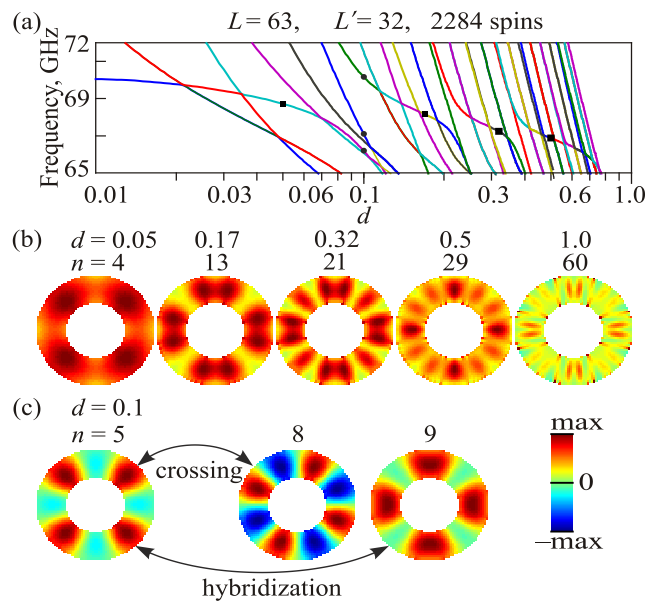

*Fig. 8.* (Color online) (a) Evolution of the fundamental mode frequencies as a function of the dipolar-to-exchange interaction ratio $d$. (b) Profiles of the fundamental mode for points marked with black squares in (a). (c) Profiles of selected modes (marked with black circles in (a)) for $d = 0.1$: hybridizing ones, fundamental and one of two (0, 4) modes, and the second (0, 4) mode which is not involved in the hybridization.

After the hybridization as $d$ continues to grow the maxima of the precession amplitude split and the symmetry of the fundamental mode is doubled ($d = 0.17$ in Fig. 8(b)). In the same time its frequency changes only little thus it crosses descending azimuthal modes (0, 5), (0, 6) and (0, 7), i.e. those which profiles do not match its own symmetry. (Modes of 5th and 7th order degenerate in pairs while modes (0, 6) are split due to the coincidence with the symmetry of the lattice.) Now the fundamental mode has 8-fold symmetry thus it matches the (0, 8) azimuthal modes. As a consequence another hybridization occurs for $d$ between 0.20 and 0.25. Again, one of these modes is ignored and the other hybridize with (0, 0) mode. Afterwards situation is repeated for $d = 0.35-0.40$ and $d = 0.55-0.57$ when fundamental mode hybridizes with one of modes (0, 12) and (0, 16), respectively. For $d > 0.75$ the mode under the question loses its fundamental character; its profile has pronounced maximums and minimums ($d = 1.0$ in Fig. 8(b)), and its frequency noticeably depends on $d$. There is no more fundamental mode in the spectrum.

## 6. Stability of the vortex configuration

According to the discussion in Sec. 3 even without performing any simulations our method gives an information about stability of the assumed magnetic configuration. It is clearly seen in Fig. 6 where in the frequency vs dipolar-to-exchange interaction ratio dependence three regions can be distinguish: two with zero-frequency modes and one without. If $d < d_1$ the exchange interaction drives the magnetic moments out of the plane of the ring. On the other hand, the dipolar interaction tends to split magnetic vortex into multi-vortex (or multi-domain) configuration, which appears for $d > d_2$. The dependence of these two critical values on the size and the shape of dots and rings was already the subject of our previous studies [23,69,73,74] so here we only recall main results.

For the dipolar-driven reorientation the dependence of $d_2$ on the size of the system is very similar in circular and square rings as well as in full circular dots. In Fig. 9 we show this dependence for rings: in (a) for circularly magnetized circular rings and in (c) for square rings in the Landau state. In figures (b) and (d) spin wave profiles of the lowest modes are shown, respectively. In both cases for full dots ($L' = 0$) the profile is strongly localized at the centre of the dot. The elimination of only four central spins destroys the central localization of the lowest mode ($L' = 2$). Now for the square rings the maximum amplitude occurs along the lines between the corners of the hole and the external edges of the ring. For the circular rings the maximum amplitude lies along high spin density lines. In both types of ring the change in the profile of the lowest mode results in a drastic change in the critical value $d_2$. Further increasing of the internal size of the ring brings continuous change of the profile which results in the smooth increasing of $d_2$. This means with increasing of the hole in the ring the stability of the vortex increases as well (the vortex in stable for weaker exchange interactions).

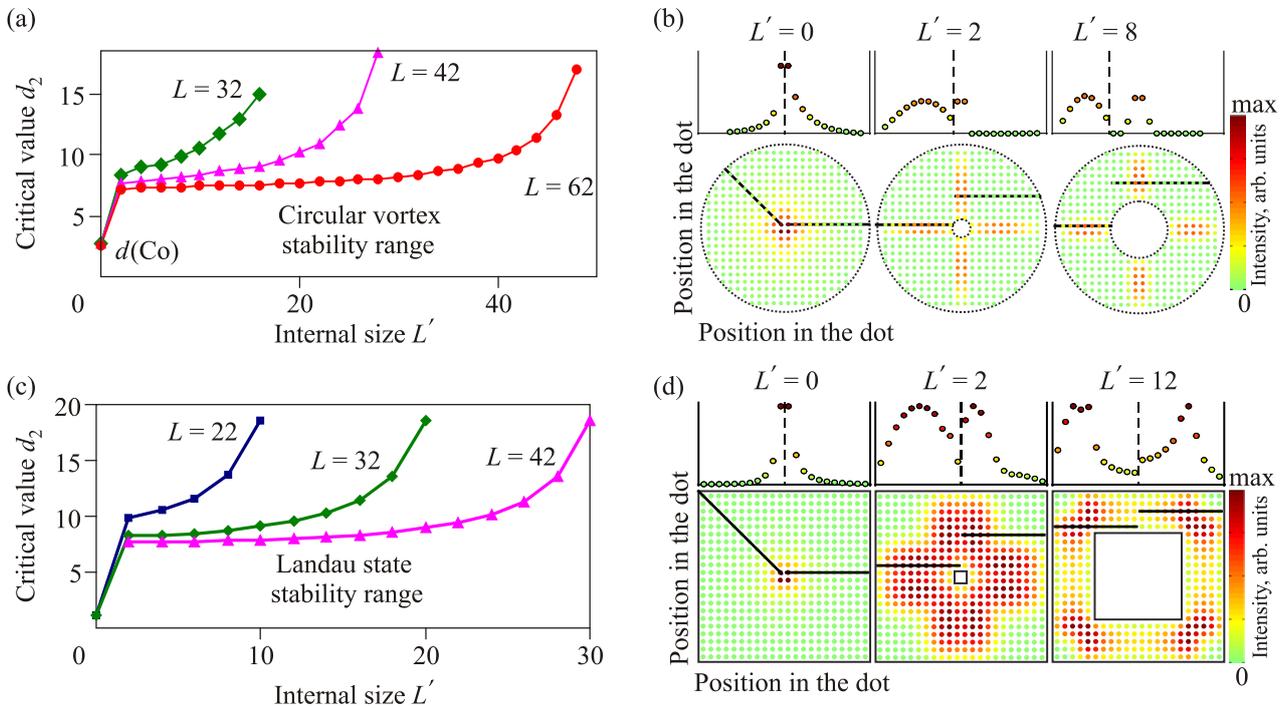

*Fig. 9.* (Color online) (a), (c) Critical value $d_2$ vs the internal size $L'$ of the (a) circular and (c) square ring for different external sizes $L$. (b), (d) Spin wave profiles of the lowest-frequency mode in the (b) circular and (d) square ring in the in-plane vortex state for $d \approx d_2$ for three internal sizes $L'$. The external sizes of rings are fixed at $L = 22$. Above each profile, its section along the indicated line. Figures (a), (b) are from [73] and (c), (d) from [74].

In Fig. 10 we show the critical $d$ for exchange driven reorientation ($d_1$) for similar rings as in Fig. 9 but now dependencies for square and circular rings are completely different. First of all, for full dots critical value of $d$ is very high, i.e. the in-plane vortex could be stable for very weak exchange interactions. For circular rings $d_1$ smoothly decreases with increasing the internal size of the ring starting from full dot. This behavior is related to the change of the profile of the lowest mode. For full dot the profile is localized at the centre. Removing of the central part of the dot results in decreasing localization at the internal edge of the ring. For larger hole in the ring the profile is almost uniform in radial direction and $d_1$ is very little dependent on the $L'$. Decreasing $d_1$ means increasing stability of the vortex, stronger exchange interactions are necessary to destroy it and for rings large enough the in-plane vortex is stable even for such strong magnetic materials as cobalt ($d_1 < d_{Co}$).

Completely different is the dependence for square dots. In this case introducing of the hole in the middle of the dot results in rapid fall of $d_1$ but afterwards its value is constant for the wide range of $L'$. Again this behavior is an effect of the change in the spin wave profile. Introducing of a hole destroy centre localization, similarly to the $d_2$ cases, but now the localization is shifted to the corners of a ring and such profile stands unchanged until the hole is sufficiently large.

In Fig. 11 we show the influence of the lattice the ring is cut from on the stability of the in-plane vortex (figure taken from Ref. 69). For both lattices, square and hexagonal, change in the size of the ring leads to the same results: $d_1$ decreases as the diameter (internal or external) increases which is related to dipolar interactions gaining of importance. The critical value of $d$ is larger for the hexagonal lattice than for the square lattice, i.e. for the hexagonal lattice the in-plane vortex is destabilized by weaker exchange interaction. This is due to the number of nearest neighbors, greater in the hexagonal lattice, implying increased impact of the exchange interaction. Also in Monte Carlo simulations similar effect was observed for full dots [65]. The feature weakens substantially with increasing of any of the diameters of the ring as a consequence of the increase in the impact of the dipolar interaction, which reduces the effect of the change in the number of nearest neighbors.

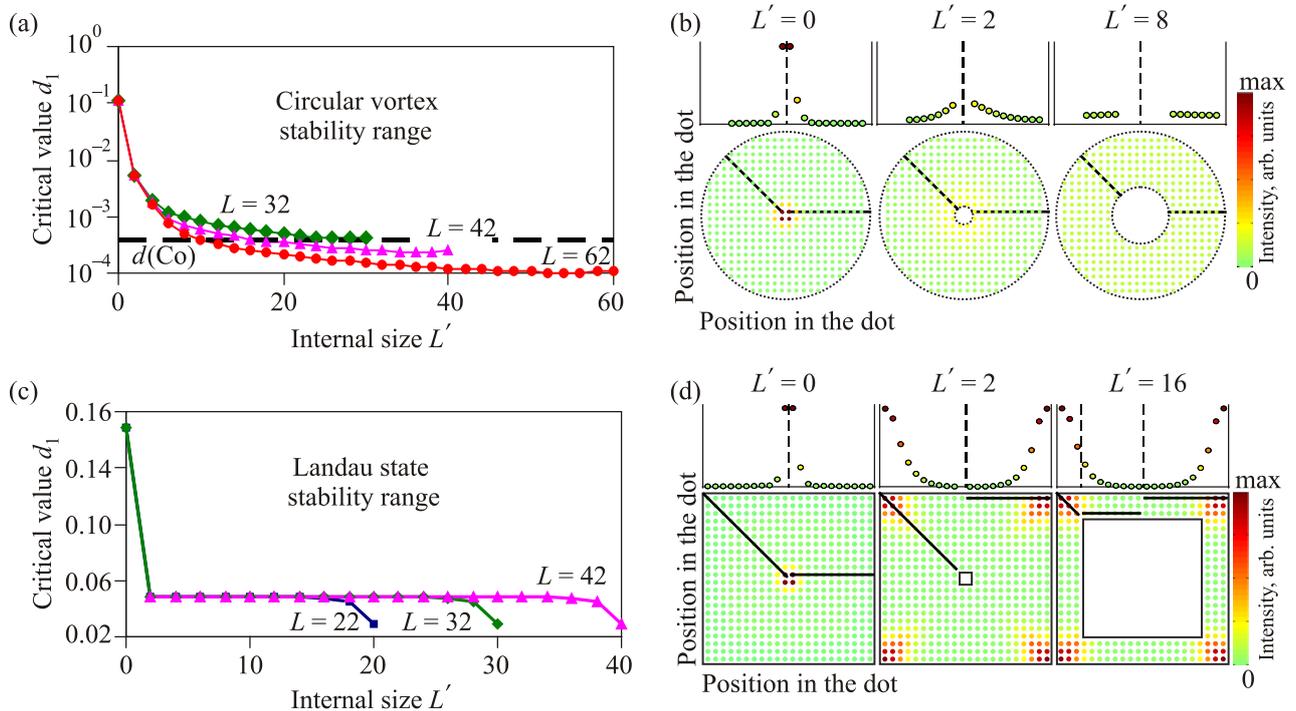

*Fig. 10.* (Color online) (a), (c) Critical value $d_1$ vs the internal size $L'$ of the (a) circular and (c) square ring for different external sizes $L$. The dashed line in (a) indicates the value of d for Co/Cu(001) calculated from experimental results reported in [66]. (b, d) Spin wave profiles of the lowest-frequency mode in the (b) circular and (d) square ring in the in-plane vortex state for $d \approx d_1$ for three internal sizes $L'$. The external sizes of rings are fixed at $L = 22$. Above each profile, its section along the indicated line. Figures (a), (b) are from [73] and (c), (d) from [74].

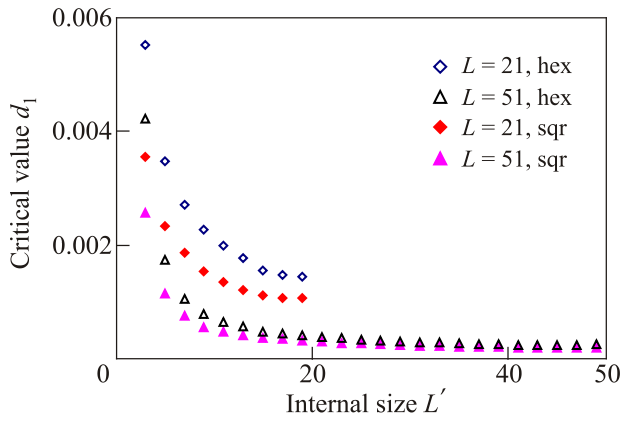

*Fig. 11.* (Color online) Critical value $d_1$ of the dipolar-to-exchange interaction ratio $d$ vs the internal size $L'$ for the circularly magnetized rings of external size $L = 21$ (diamonds) and $L = 51$ (triangles). Empty and full marks refer to rings based on the hexagonal and square lattice, respectively. (Figure taken from Ref. 69.)

## 7. Conclusion

In this paper we have shown detailed derivation of the theoretical method, based on the discrete version of the damping-free Landau–Lifshitz equation, useful to study confined systems cut out from a discrete lattice of magnetic moments. The method is general enough to be used for the system of an arbitrary size or shape as well as the magnetic configuration and the lattice the system is based on. If exchange interactions are omitted the method is applicable also for nonperiodic systems. Two main adventages of this approach are the spin wave spectrum obtained directly form diagonalization of the dynamical matrix and very short time of calculations in comparison with time-domain simulations. On the other hand, the lack of simulations is the weakest point of the method: it is useful for very simple magnetic configurations such as uniform magnetization or in-plane vortices. The method can be used for more complicated configurations as well but for magnetic state simulated rather then assumed. In such case the total time of calculation is similar to simulation methods but the frequencies and profiles of the spin waves are obtained directly from the diagonalization without the usage of the Fourier transformation. For the dipolar-exchange case we introduce the dipolar-to-exchange interaction ratio $d$ and study how this parameter, the only parameter of the model, influences the spin wave spectrum of the system.

As an example we use a circularly magnetized ring cut from a square lattice. We show that internal structure of the discrete lattice has a significant impact on the spin wave spectrum. The symmetry of the lattice cause frequency splitting of the pair of azimuthal modes if it match the symmetry of the mode profiles. For the square lattice the frequencies of modes with even azimuthal number are split while in the case of hexagonal lattice for azimuthal number divisible by 3. The symmetry of the lattice is reflected also in the non-uniformity of the fundamental mode profile. This effect has two consequences: there is a weak dependence of the fundamental mode frequency on $d$ and the hybridization with the azimuthal mode occurs if its profile meets two conditions. The first one is the same symmetry of the profiles of both hybridizing modes. The second one is the matching of the in-phase anti-nodes in the azimuthal mode with the maximums in the fundamental mode.

We also elucidate the influence of the competition between dipolar and exchange interactions of the spin wave spectrum. We found that not only stable magnetic configuration but also the lowest-frequency mode is indicative of this competition. Far from the critical values of $d$ exchange interactions prefer lower azimuthal number while dipolar interactions favour higher ones, regardless of whether their predomination is due to the material or size of the dot. The profile of the lowest mode carry information on the stability of the magnetic configuration as well. The lowest mode is quasi-uniform close to the exchange driven reorientation or localized along the high spin density lines close to the dipolar one.

We also provide a selected review concerning the spin wave excitations in full circular dots with a special attention paid on the lowest-frequency mode. In this case this mode is localized at the vortex centre if it is a soft mode, i.e., one type of interaction prevails ($d$ close to one of its critical values). Otherwise the lowest mode is an azimuthal mode of different order and reflects the competition between the dipolar and exchange interactions, similarly to the rings. These findings help us to explain the diversity of the results reported in numerous experimental and micromagnetic studies.

This project has received funding from the European Union's Horizon 2020 research and innovation programme under the Marie Skłodowska-Curie grant agreement No. 644348.

1. H. Puszkarski, J.-C.S. Lévy, and S. Mamica, *Phys. Lett. A* **246**, 347 (1998).
2. S. Mamica, R. Józefowicz, and H. Puszkarski, *Acta Phys. Polon. A* **94**, 79 (1998).
3. S. Mamica, H. Puszkarski, and J.C.S. Lévy, *Phys. Status Solidi B* **218**, 561 (2000).
4. M. Krawczyk, S. Mamica, J.W. Kłos, J. Romero-Vivas, M. Mruczkiewicz, and A. Barman, *J. Appl. Phys.* **109**, 113903 (2011).
5. S. Pal, B. Rana, S. Saha, R. Mandal, O. Hellwig, J. Romero-Vivas, S. Mamica, J.W. Klos, M. Mruczkiewicz, M.L. Sokolovskyy, M. Krawczyk, and A. Barman, *J. Appl. Phys.* **111**, 07C 507 (2012).
6. M. Krawczyk, H. Puszkarski, J.-C. Lévy, S. Mamica, and D. Mercier, *J. Magn. Magn. Mater.* **246**, 93 (2002).
7. J.W. Kłos, M.L. Sokolovskyy, S. Mamica, and M. Krawczyk, *J. Appl. Phys.* **111**, 123910 (2012).

8. S. Mamica, M. Krawczyk, and J.W. Kłos, *Adv. Cond. Mat. Phys*. **2012**, 161387 (2012).
9. S. Mamica, M. Krawczyk, M.L. Sokolovskyy, and J. Romero-Vivas, *Phys. Rev. B* **86**, 144402 (2012).
10. J. Romero Vivas, S. Mamica, M. Krawczyk, and V.V. Kruglyak, *Phys. Rev. B* **86**, 144417 (2012).
11. M.F. Lai, Z.H. Wei, C.R. Chang, N.A. Usov, J.C. Wu, and J.Y. Lai, *J. Magn. Magn. Mater.* **272-276**, e1331 (2004).
12. C.A.F. Vaz, M. Kläui, J.A.C. Bland, L.J. Heyderman, C. David, and F. Nolting, *Nucl. Instrum. Methods B* **246**, 13 (2006).
13. K.L. Metlov and Y.P. Lee, *Appl. Phys. Lett.* **92**, 112506 (2008).
14. W. Zhang, R. Singh, N. Bray-Ali, and S. Haas, *Phys. Rev. B* **77**, 144428 (2008).
15. S.-H. Chung, R.D. McMichael, D.T. Pierce, and J. Unguris, *Phys. Rev. B* **81**, 024410 (2010).
16. J.P. Park, P. Eames, D.M. Engebretson, J. Berezovsky, and P.A. Crowell, *Phys. Rev. B* **67**, 020403(R) (2003).
17. L. Rondin, J.-P. Tetienne, S. Rohart, A. Thiaville, T. Hingant, P. Spinicelli, J.-F. Roch, and V. Jacques, *Nature Commun.* **4**, 2279 (2013).
18. Ph. Depondt, J.-C.S. Lévy, and S. Mamica, *J. Phys. Cond. Matter* **25**, 466001 (2013).
19. T. Shinjo, T. Okuno, R. Hassdorf, K. Shigeto, and T. Ono, *Science* **289**, 930 (2000).
20. A. Wachowiak, J. Wiebe, M. Bode, O. Pietzsch, M. Morgenstern, and R. Wiesendanger, *Science* **298**, 577 (2002).
21. J. Miltat and A. Thiaville, *Science* **298**, 555 (2002).
22. S.P. Li, D. Peyrade, M. Natali, A. Lebib, Y. Chen, U. Ebels, L.D. Buda, and K. Ounadjela, *Phys. Rev. Lett.* **86**, 1102 (2001).
23. S. Mamica, J.-C.S. Lévy, Ph. Depondt, M. Krawczyk, *J. Nanopart. Res*. **13**, 6075 (2011).
24. V.S. Pribiag, I.N. Krivorotov, G.D. Fuchs, P.M. Braganca, O. Ozatay, J.C. Sankey, D.C. Ralph, and R.A. Buhrman, *Nat. Phys.* **3**, 498 (2007).
25. K.Y. Guslienko and J. Spintron, *Magn. Nanomater.* **1**, 70 (2012).
26. V.E. Demidov, H. Ulrichs, S. Urazhdin, S.O. Demokritov, V. Bessonov, R. Gieniusz, and A. Maziewski, *Appl. Phys. Lett.* **99**, 012505 (2011).
27. R.P. Cowburn and M.E. Welland, *Science* **287**, 1466 (2000).
28. C.A. Ross, *Annu. Rev. Mater. Res.* **31**, 203 (2001).
29. M.R. Mozaffari and K. Esfarjani, *Physica B* **399**, 81 (2007).
30. R. Zivieri and F. Nizzoli, *Phys. Rev. B* **78**, 064418 (2008).
31. S. Rohart, P. Campiglio, V. Repain, Y. Nahas, C. Chacon, Y. Girard, J. Lagoute, A. Thiaville, and S. Rousset, *Phys. Rev. Lett.* **104**, 137202 (2010).
32. M. Buess, R. Höllinger, T. Haug, K. Perzlmaier, U. Krey, D. Pescia, M.R. Scheinfein, D. Weiss, and C.H. Back, *Phys. Rev. Lett.* **93**, 077207 (2004).
33. J.P. Park and P.A. Crowell, *Phys. Rev. Lett.* **95**, 167201 (2005).
34. V.E. Demidov, U.-H. Hansen, and S.O. Demokritov, *Phys. Rev. Lett.* **98**, 157203 (2007).
35. F.G. Aliev, J.F. Sierra, A.A. Awad, G.N. Kakazei, D.-S. Han, S.-K. Kim, V. Metlushko, B. Ilic, and K.Y. Guslienko, *Phys. Rev. B* **79**, 174433 (2009).
36. V.V. Kruglyak, P.S. Keatley, A. Neudert, R.J. Hicken, J.R. Childress, and J.A. Katine, *Phys. Rev. Lett.* **104**, 027201 (2010).
37. K. Vogt, O. Sukhostavets, H. Schultheiss, B. Obry, P. Pirro, A.A. Serga, T. Sebastian, J. Gonzalez, K.Y. Guslienko, and B. Hillebrands, *Phys. Rev. B* **84**, 174401 (2011).
38. G.M. Wysin and A.R. Völkel, *Phys. Rev. B* **52**, 7412 (1995).
39. M. Grimsditch, L. Giovannini, F. Montoncello, F. Nizzoli, G.K. Leaf, and H.G. Kaper, *Phys. Rev. B* **70**, 054409 (2004).
40. B.A. Ivanov and C.E. Zaspel, *Phys. Rev. Lett.* **94**, 027205 (2005).
41. C.E. Zaspel, B.A. Ivanov, J.P. Park, and P.A. Crowell, *Phys. Rev. B* **72**, 024427 (2005).
42. K. Rivkin, L.E. DeLong, and J.B. Ketterson, *J. Appl. Phys.* **97**, 10E309 (2005).
43. R. Zivieri and R.L. Stamps, *Phys. Rev. B* **73**, 144422 (2006).
44. E.Yu. Vedmedenko, H.P. Oepen, A. Ghazali, J.-C.S. Lévy, and J. Kirschner, *Phys. Rev. Lett.* **84**, 5884 (2000).
45. V. Novosad, M. Grimsditch, K.Yu. Guslienko, P. Vavassori, Y. Otani, and S.D. Bader, *Phys. Rev. B* **66**, 052407 (2002).
46. B.V. Costa, *Braz. J. Phys.* **41**, 94 (2011).
47. M. Dvornik, P.V. Bondarenko, B.A. Ivanov, and V.V. Kruglyak, *J. Appl. Phys.* **109**, 07B912 (2011).
48. D. Toscano, S.A. Leonel, P.Z. Coura, F. Sato, R.A. Dias, and B.V. Costa, *Appl. Phys. Lett.* **101**, 252402 (2012).
49. Ph. Depondt and J.-C.S. Lévy, *Phys. Lett. A* **376**, 3411 (2012).
50. C.E. Zaspel and B.A. Ivanov, *J. Magn. Magn. Mater.* **286**, 366 (2005).
51. J. Podbielski, F. Giesen, M. Berginski, N. Hoyer, and D. Grundler, *Superlatt. Microstruct.* **37**, 341 (2005).
52. C.G. Tan, H.S. Lim, Z.K. Wang, S.C. Ng, M.H. Kuok, S. Goolaup, A.O. Adeyeye, and N. Singh, *J. Magn. Magn. Mater.* **320**, 475 (2008).
53. M. Buess, T.P.J. Knowles, R. Hollinger, T. Haug, U. Krey, D. Weiss, D. Pescia, M.R. Scheinfein, and C.H. Back, *Phys. Rev. B* **71**, 104415 (2005).
54. O.G. Heinonen, D.K. Schreiber, and A.K. Petford-Long, *Phys. Rev. B* **76**, 144407 (2007).
55. I. Neudecker, F. Hoffmann, G. Woltersdorf, and C.H. Back, *J. Phys. D* **41**, 164010 (2008).
56. R. Wang and X. Dong, *Appl. Phys. Lett.* **100**, 082402 (2012).
57. X. Zhu, Z. Liu, V. Metlushko, P. Grütter, and M.R. Freeman, *Phys. Rev. B* **71**, 180408(R) (2005).
58. F. Boust, N. Vukadinovic, and S. Labbé, *J. Magn. Magn. Mater.* **272–276**, 708 (2004).
59. Z. Liu, R.D. Sydora, and M.R. Freeman, *Phys. Rev. B* **77**, 174410 (2008).
60. S. Mamica, J.-C.S. Lévy, and M. Krawczyk, *J. Phys. D* **47**, 015003 (2014).

61. K.Yu. Guslienko, W. Scholz, R.W. Chantrell, and V. Novosad, *Phys. Rev. B* **71**, 144407 (2005).
62. R. Zivieri and F. Nizzoli, *Phys. Rev. B* **71**, 014411 (2005).
63. N.A. Usov and S.E. Peschany, *J. Magn. Magn. Mater.* **118**, L290 (1993).
64. R. Skomski, *Simple Models of Magnetism*, Oxford University Press, Oxford (2008).
65. J.C.S. Rocha, P.Z. Coura, S.A. Leonel, R.A. Dias, and B.V. Costa, *J. Appl. Phys.* **107**, 053903 (2010).
66. R. Vollmer, M. Etzkorn, P.S.A. Kumar, H. Ibach, and J. Kirschner, *J. Appl. Phys.* **95**, 7435 (2004).
67. F. Hoffmann, G. Woltersdorf, K. Perzlmaier, A.N. Slavin, V.S. Tiberkevich, A. Bischof, D. Weiss, and C.H. Back, *Phys. Rev. B* **76**, 014416 (2007).
68. K.Y. Guslienko, A.N. Slavin, V. Tiberkevich, and S.-K. Kim, *Phys. Rev. Lett.* **101**, 247203 (2008).
69. S. Mamica, *J. Appl. Phys.* **114**, 233906 (2013).
70. L. Giovannini, F. Montoncello, R. Zivieri, and F. Nizzoli, *J. Phys.: Condens. Matter* **19**, 225008 (2007).
71. F. Montoncello, L. Giovannini, F. Nizzoli, R. Zivieri, G. Consolo, and G. Gubbiotti, *J. Magn. Magn. Mater.* **322**, 2330 (2010).
72. T. Fischbacher, M. Franchin, G. Bordignon, and H. Fangohr, *IEEE Transact. Magnet.* **43**, 2896 (2007).
73. S. Mamica, *J. Appl. Phys.* **113**, 093901 (2013).
74. S. Mamica, J.-C.S. Lévy, M. Krawczyk, and Ph. Depondt, *J. Appl. Phys.* **112**, 043901 (2012).